# Controlled Curie temperature, magneto-crystal anisotropy, and valley polarization in 2D ferromagnetic Janus 2H-VSeS monolayer

Cunquan Li[1] and Yukai An[1,*]

[1] *Key Laboratory of Display Materials and Photoelectric Devices, Ministry of Education, Tianjin Key Laboratory for Photoelectric Materials and Devices, National Demonstration Center for Experimental Function Materials Education, School of Material Science and Engineering, Tianjin University of Technology, Tianjin, 300384, China*

Inspired by the successful synthesis of two-dimensional (2D) V-based Janus dichloride monolayers with intrinsic ferromagnetism and high Curie temperature ($T_c$), the electronic structure, spin-valley splitting and magnetic anisotropy of Janus 2H-VSeS monolayers are investigated in detailed using first-principles calculations. The results show that the Janus 2H-VSeS monolayer exhibits a large valley splitting of 105 meV, high $T_c$ of 278 K and good magnetocrystalline anisotropy (0.31 meV) contributed by the in-plane $d_{x^2-y^2}/d_{xy}$ orbitals of V atoms. The biaxial strain ($-8\%<\varepsilon<8\%$) can effectively tune the magnetic moments of V atom, valley splitting $\Delta$E, $T_c$ and MAE of Janus 2H-VSeS monolayer. The corresponding $\Delta$E and $T_c$ are adjusted from 72 meV to 106.8 meV and from 180 K to 340 K, respectively. The electronic phase transition from bipolar magnetic semiconductor (BMS) to half-semiconductor (HSC), spin gapless semiconductor (SGS), and half-metal (HM) is also observed due to the change of V 3d-orbital occupation. Due to the broken space- and time-reversal symmetry, the opposite valley charge carriers carry opposite Berry curvature, which leads to prominent anomalous Hall conductivity at the K and K′ valleys. The maximum modulation of Berry curvature can reach to 45% and 9.5% by applying the biaxial strain and charge carrier doping, respectively. The stable in-plane magnetocrystalline anisotropy and large spontaneous valley polarization make the ferromagnetic Janus 2H-VSeS monolayer a promising material for achieving the spintronics and valleytronics devices.

## I. INTRODUCTION

Since the successful discovery and synthesis of graphene, [1] two-dimensional (2D) materials with hexagonal lattice structures have attracted a great deal of interest due to their excellent physical and chemical properties compared with the bulk counterparts. [2] As the members of 2D family, transition metal dichalcogenides (TMDs) $MX_2$ (M=Mo, W; X=S, Se) with a central M sublayer sandwiched by two mirror-symmetric X sublayers possess a tunable band gap (1-2 eV) and become a good candidate for optoelectronic, valley electronic and nanoelectronic devices. [3, 4] Due to the lack of inversion symmetry, TMDs possess a pair of inequivalent degeneracy energy valleys at the K and K′ points in momentum space and exhibit extraordinary quantum effects, such as valley-spin locking and valley-spin Hall effects. [5–7] The energy valley can be utilized and manipulated, making it an excellent information carrier for charge and spin to store information and perform logic operations. The biggest challenge in the development of valleytronics is to achieve the manipulation and generation of valley splitting by lifting the degeneracy between the K and K′ valleys. Until now, multiple pathways to lift the valley degeneracy have been performed, including the circularly polarized light pumping, [8] external magnetic fields, [9] transition-metal (TM) atoms doping or adsorption [10] and magnetic proximity effect. [11] However, the generated valley splitting will disappear and return to the intrinsic paravalley states once the external field is removed, which is impracticable for the application in valleytronics devices. Therefore, searching for 2D materials with intrinsic valley splitting is crucial. Recently, the 2D ferrovalley monolayers ($VS_2$, [12] $VSe_2$ [13] and $VTe_2$ [14]) with intrinsic ferromagnetic ordering and spontaneous valley splitting have been successfully synthesized and attracted considerable attention due to their strong magneto-valley coupling effects. The spontaneous valley splitting in the V-based TMDs monolayers originates from the magnetic interaction between the V 3d electrons, which does not depend on the external fields and can achieve simultaneous manipulation of spin and valley degrees of freedom.

Stable Janus group VI chalcogenides MXY (M=Mo, W; X, Y=S, Se, Te; X≠Y) are firstly proposed by Cheng et al., [15] where the Janus MoSSe have been successfully synthesized via experiments using the modified chemical vapor deposition methods. [16] At the same time, some Janus systems, including Janus graphene, [17] Janus graphene oxide, [18] and group-III chalcogenides $NN'X_2$ (N, N′=Ga, In; N≠N′; X=S, Se, Te) [19] monolayers are also systematically investigated, which exhibit extraordinary physical properties with tunable electric dipole and high carrier mobility. Additionally, 2D Janus TMDs with intrinsic ferromagnetism, such as Janus MnXY (X, Y=S, Se, Te; X≠Y) [20, 21] and $Cr(I, X)_3$ (X=Cl, I and Br) [22] monolayers have been proved to exhibit a large Dzyaloshinskii-Moriya interaction behavior. [23] The ex-perimental realization of 2D ferrovalley materials, especially for the 2H-$VSe_2$ monolayer, may provide an exciting new platform for the Janus structure with intrinsic ferromagnetic semiconductors. It is expected that the ordered Janus 2H-VSeS monolayer can be grown by a similar modified CVD method as described above and introduce new physical phenomena. In addition, the Janus 2H-VSeS monolayer are firstly predicted by Du et al [24] and they comprehensively investigated its piezoelectricity and ferroelasticity properties. Further, it is significant to gain greater insight into the intrinsic physical properties of Janus 2H-VSeS monolayer.

* ykan@tjut.edu.cn

In this work, the geometric structure, kinetic stability, Curie temperature ($T_c$), magnetocrystalline anisotropy and valley splitting of Janus 2H-VSeS monolayers under various biaxial strains and charge carrier doping are systematically discussed by the first-principles calculations. The results show that the electronic structures of Janus 2H-VSeS monolayer can be effectively modulated from bipolar magnetic semiconductor (BMS) to semiconductor (HSC), spin-gap-free semiconductor (SGS) and half-metal (HM), which results an obvious change in the magnetic moments of V atoms, valley splitting, band gap and $T_c$. Furthermore, the Berry curvatures at the K- and K′-valleys show the opposite signs due to the break of mirror symmetry along with the out-of-plane, which can also be effectively modulated by the biaxial strain and charge carrier doping. These results provide new insights for adjusting the electronic structure and magnetic properties of Janus 2H-VSeS monolayer, which is very helpful for the design of new nanoelectronic devices.

## II. COMPUTATIONAL METHODS

The first-principles calculations are performed using the plane-wave basis Vienna Ab initio Simulation Package (VASP), [25–28] based on the density functional theory (DFT). The generalized gradient approximation (GGA) functional in Perdew-Burke-Ernzerhof (PBE) is adopted. [29, 30] Considering the strong-correlated correction of V 3d localized electrons, the LSDA+U method is adopted, [31] where the effective onsite Coulomb interaction parameter (U=2.00 eV) and exchange interaction parameter (J=0.87 eV) have been added to it. [32, 33] A vacuum space of 18 Å in the z-direction is set to avoid the adjacent interactions. The cut-off energy is set at 500 eV and the criteria of energy and atom force convergence are $10^{-6}$ Å and $10^{-3}$ eV/Å$^{-1}$, respectively. The Brillouin zone is sampled with 16×16×1 Γ-centered Monkhorst-Pack grids. [34] The effect of van der Waals (vdW) corrections (DFT-D3 method) [35] are considered during the structure optimizations and the VASPKIT [14] code is used to process the VASP data. The phonon spectrum and phonon density of states are calculated by the PHONOPY [36] based on density-functional perturbation theory (DFPT) and a 4×4×1 supercell is used to calculate the Hessian matrix. The spin-orbit coupling (SOC) and non-collinear magnetism are considered in the calculations of electronic structure. For the calculation of Berry curvature and anomalous Hall conductivity, the maximally localized Wannier functions (MLWFs) are constructed using the WANNIER90 package. [37, 38] In the case of Janus 2H-VSeS monolayer, there are 22 bands in the energy range from about −6 to 4 eV, mainly formed by V d orbitals and X (X=Se, S) p orbitals. Ten d orbitals on atom V and six p orbitals on each atom X are chosen as the initial guess of Wannier functions. After less than 400 iterative steps, the total Wannier spreads are well converged down to $10^{-6}$ Å$^2$. A fine k-mesh of 30×30×1 is used in WANNIER interpolation. The Monte Carlo simulation procedure MCSOLVER [39, 40] based on the Wolff algorithm of the classical Heisenberg model is used to estimate the $T_c$. In the specified temperature interval, we completely thermalize the system to equilibrium with 120,000 scans starting from the ferromagnetic order, and all statistics are obtained from the next 720,000 scans.

## III. RESULTS AND DISCUSSION

It has been reported that the trigonal prismatic (2H) phase is more energetically stable than the octahedral (1T) phase in the Janus VSeS monolayer. [24] Thus, the geometry of Janus 2H-VSeS monolayer with a S-V-Se sandwich structure is fabricated by MedeA-VASP materials design software, as shown in Fig. 1(a). After the optimization of geometry, the corresponding V-Se and V-S bond lengths are 2.51 and 2.36 Å, respectively. The optimized lattice constant is 3.259 Å, which is between those of 2H-$VS_2$ (3.18 Å) and 2H-$VSe_2$ (3.34 Å) monolayers. [13, 41] For the optimized geometry, the 2H-$VSe_2$ and 2H-$VS_2$ monolayers belong to the $D_{3h}$ point group, while the Janus 2H-VSeS monolayer breaks the off-plane structure symmetry and presents the $C_{3v}$ point group. The calculated average electrostatic potential along z-axis is rather asymmetric, as shown in Fig. 1(b). Obviously, the electrostatic potential of the S side is smaller than that of the Se side, namely $V_S<V_{Se}$, implying a larger electronegativity of the S atom. Each V atom contributes a magnetic moment of 0.978 $\mu_B$ in the Janus 2H-VSeS monolayer. Meanwhile, the difference of electronegativity between the S and Se atoms can alter the local magnetic interaction, which induces a larger magnetic moment for the V atom. To confirm the magnetic ground states of Janus 2H-VSeS monolayer, the ferromagnetic (FM) and antiferromagnetic (AFM) states in a 2×2×1 supercell are calculated. The total magnetic moment is 4.0 $\mu_B$ for the Janus 2H-VSeS monolayer with FM state, while the total magnetic moment is 0 $\mu_B$ with AFM state. The ferromagnetic stability energy ($\Delta E=E_{FM-AFM}$) of 2×2×1 supercell is −260 meV<0, strongly suggesting that the Janus 2H-VSeS monolayer is an intrinsic ferromagnetic material. As is known, the Janus 2H-VSeS monolayer suffers from the spin-wave, so the non-magnetic (NM) ground state is a supercell with Peierls distortion, and the phonon dispersion is also considered to be unstable due to the presence of imaginary frequencies in the AFM state. Therefore, to estimate the stability of Janus 2H-VSeS monolayer, the phonon spectrum and phonon density of states with FM state are finally calculated in a 4×4×1 supercell. As shown in Fig. 1(c), the absence of significant imaginary phonon modes over the whole Brillouin zone strongly suggests that the Janus 2H-VSeS monolayer is dynamically stable, which is consistent with the previous report. [24] The highest frequency is calculated to be 12.35 Thz, which is between the reported values for the 2H-$VSe_2$ (10.02 Thz) and 2H-$VS_2$ (13.7 Thz) monolayers. [41–43] The contribution at low frequencies ($0<\varepsilon<5$ THz) mainly comes from the Se atoms, in contrast to the V and S atoms which make the main contribution at high frequencies ($6<\varepsilon<12$ THz), as shown in Fig. S1. It is maybe due to that the V (atomic weight: 50.94) and S (32.066) atoms are lighter compared to the Se (78.96) atoms. This can also remarkably affect the thermal properties of Janus 2H-VSeS monolayer. At low temperature, the thermal conductivity is mainly contributed by the phonons of low frequency Se atom. However, at high temperature, the thermal conductivity is mainly contributed by phonons of higher frequency V/S atoms. The spin-polarized band structure of Janus 2H-VSeS monolayer is shown in Fig. 1(d). One can see that the valence band maximum (VBM) occupied by spin-up ($\alpha$) electrons is located at the Γ point of Brillouin zone, while the conduction band maximum (CBM) contributed by spin-down ($\beta$) electrons is located at the M point. Thus, the Janus 2H-VSeS monolayer is a bipolar magnetic semiconductor (BMS) with an indirect energy gap of 0.407 eV. Clearly, there exists a Dirac valley contributed by the spin-$\alpha$ states at the K and K′ points, as shown by the two black lines closest to the valley gap in Fig. 1(d). The two K and K′ valleys are degenerate in energy with identical valley gaps of 0.714 eV. The band structures of Janus 2H-VSeS monolayer with considering SOC is shown in Fig. 1(e). It is obvious that the valley energies at the K and K′ points in the conduction band remain close, while the energy at the K′ point is higher than that at the K point in the valence band. Hence, the valley degeneracy is broken, inducing a large valley splitting (ΔE=ΔK−ΔK′) of 105 meV, which is usually manifested as two split peaks in the photoluminescence spectrum. Fig. 1(f) shows the V d orbital-resolved band structure of Janus 2H-VSeS monolayer with considering SOC. Both the VBM and CBM at the K and K′ points are mainly contributed by V atoms. At the two unequal K and K′ valleys, the CBM is mainly contributed by the in-plane V $d_{x^2-y^2}/d_{xy}$ orbitals, while the VBM at the K and K′ points is dominated by the V $d_{z^2}$ orbital. Fig. S2(a) shows the corresponding 3D band structure of Janus 2H-VSeS monolayer with considering SOC. It is obvious that the energy of VBM at the Γ point is the global maximum, while that at the K and K′ points is only a local maximum. At the same time, the energy of CBM at the K and K′ points is also not globally minimal, and the global maximum is at the M point. 2D projected band structure of valence band and conduction band at the $k_x k_y$-plane is shown in Fig. S2(b)-(c). Clearly, the six valleys within the Brillouin zone degenerate in energy with respect to each other. This reflects that the SOC effect achieves large valley polarization in the Janus 2H-VSeS monolayer.

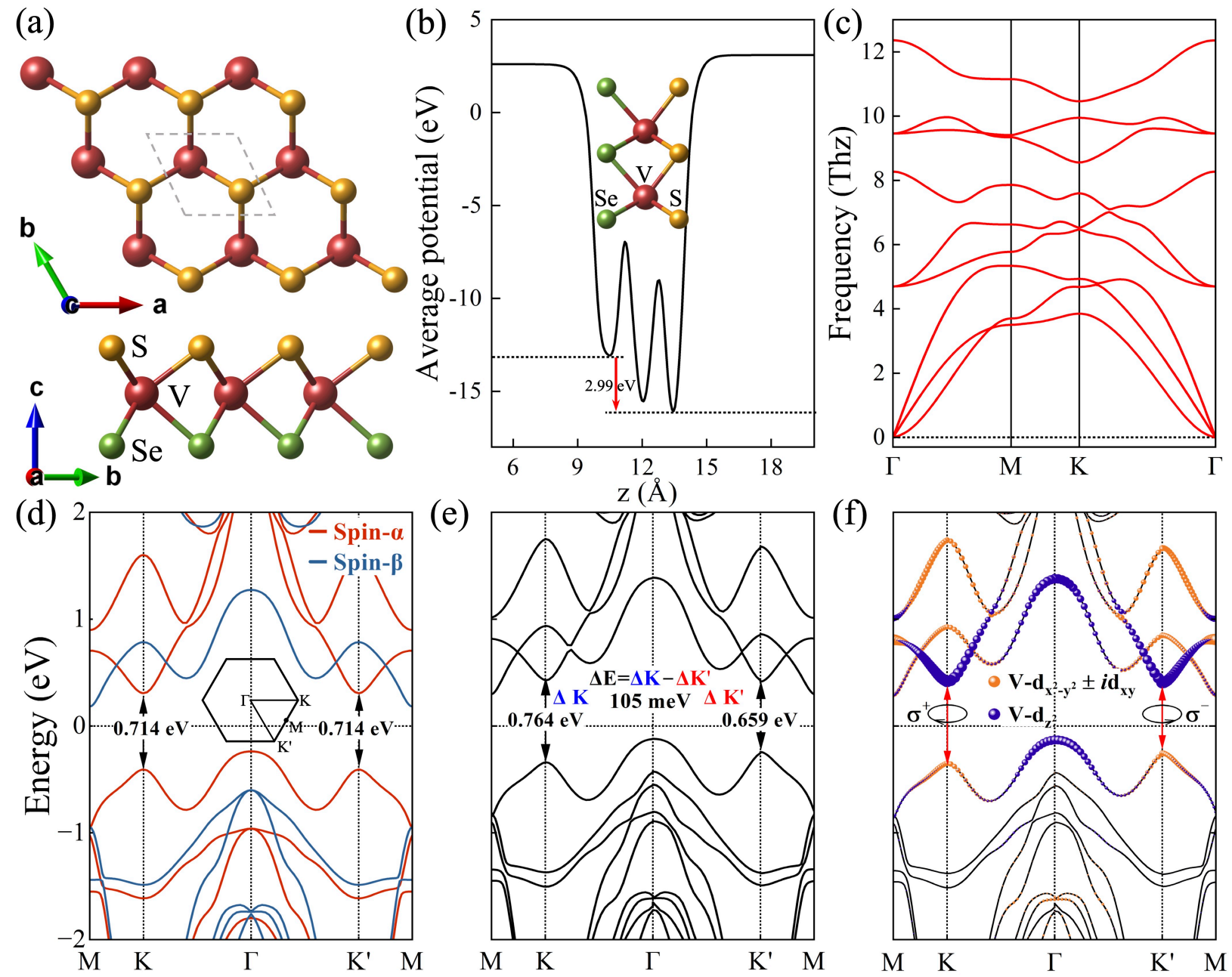

FIG. 1. (a) Top and side views of Janus 2H-VSeS monolayer. The red, yellow, and green spheres represent V, S and Se atoms, respectively. (b) The average electric potential along z direction, (c) phonon spectrum for the Janus 2H-VSeS monolayer. (d) The spin-polarized band structures of Janus 2H-VSeS monolayer. The red and blue lines represent the spin-up($\alpha$) and spin-down($\beta$) band structures. The insert is the first Brillouin zone containing the K and K′ points. (e) The band structures of Janus 2H-VSeS monolayer with considering SOC. (f) The V d orbital-resolved band structure of Janus 2H-VSeS monolayer with considering SOC, the purple and orange triangles denote the contribution from the $d_{z^2}$ and $d_{x^2-y^2}/d_{xy}$ orbitals of V atom, respectively. The arrows between the VBM and the CBM denote the valley-selective optical transitions induced by left and right circularly polarized light $\sigma^+$ and $\sigma^-$.

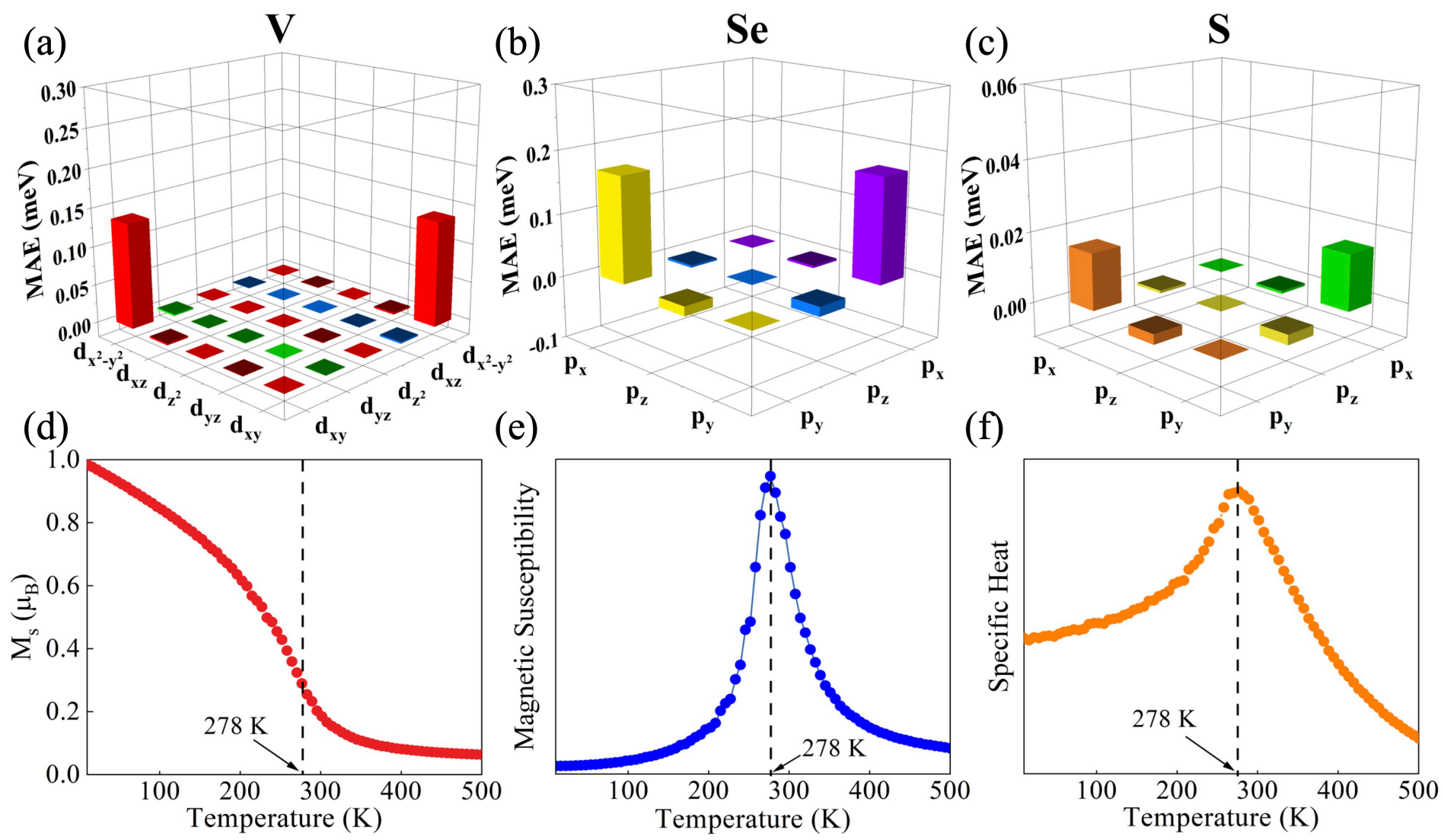

FIG. 2. (a)-(c) The V-d and Se(S)-p orbital-resolved MAEs of Janus 2H-VSeS monolayer. Dependence of (d) magnetic moment, (e) magnetic susceptibility, and (f) specific heat capacity on temperature for the Janus 2H-VSeS monolayer.

Considering the magnetic moment switching from the in-plane [100] axis to the out-of-plane [001] axis, the energy it spends is called magneto-crystal anisotropy energy (MAE). MAE($E_{[001]} - E_{[100]}$) can be defined by the following equation:

$$\mathrm{MAE} = \xi^2 \sum_{o.u} \left( \left|\langle o \left| L_x \right| u \rangle\right|^2 - \left|\langle o \left| L_z \right| u \rangle\right|^2 \right) / (E_u - E_o) \tag{1}$$

where the $\xi$, $o$ and $\mu$ are the SOC constants, and the energies of the occupied ($E_o$) and non-occupied ($E_u$) states, respectively. $E_{[001]}$ and $E_{[100]}$ are the energies of the magnetization in the [001] and [100] directions, respectively.

So, the positive (negative) MAE value implies the in-plane (out-of-plane) easy-axis. One can see from Eqn(1) that the orbital matrix element differences and energy differences together affect the MAE behavior, which can be used to analyze the contribution of different inter-orbit hybridization to the MAE. The V-d and Se(S)-p orbital resolved MAEs for the Janus 2H-VSeS monolayer are shown in Figs. 2(a)-(c). The calculated total MAE value is 0.31 meV and the corresponding MAE values contributed by the V-d, Se-p and S-p orbitals are 0.14, 0.15 and 0.01 meV, respectively. These strongly suggest that the Janus 2H-VSeS monolayer processes the in-plane magnetic anisotropy (IMA), which is mainly contributed by the hybridized V $d_{x^2-y^2}$ and $d_{xy}$ orbitals as well as the Se(S) $p_x$ and $p_y$ orbitals.

The $T_c$ of ferromagnetic materials can be well calculated using the classical Heisenberg Monte Carlo (MC) model. In the calculation, the spin Hamiltonian including the single-ion anisotropy can be described as following:

$$H = -\frac{1}{2} \sum_{(i,j)} \left[ J^x S_i^x S_j^x + J^y S_i^y S_j^y + J^z S_i^z S_j^z \right] - \sum_i D \left( S_i^z \right)^2 \tag{2}$$

For the entire V-atom lattice, $(i, j)$ denote the individual V-atoms in the immediate vicinity, and $i$ traverses the entire lattice. $J$ and $D$ are the spin-exchange parameter and magnetic anisotropy energy, respectively. $J$>0 favors the ferromagnetic interactions and $D$>0 favors the out-of plane easy axis. In the 4×4×1 supercell, the energies of FM and AFM states and the magnetic ion anisotropy parameter $D$ with considering SOC effect can be calculated as following:

$$\mathrm{E_{FM}} = \mathrm{E_0} - \left( \frac{1}{2} \times 6 \times 4 \right) J |S|^2 \tag{3}$$

$$\mathrm{E_{AFM}} = \mathrm{E_0} + 4 \times \left( -\frac{1}{2} \times 2 + \frac{1}{2} \times 4 \right) J |S|^2 \tag{4}$$

$$J = \frac{\mathrm{E_{AFM}} - \mathrm{E_{FM}}}{16|S|^2} \tag{5}$$

$$D = \frac{\mathrm{E_{FM}^{\alpha}} - \mathrm{E_{FM}^{z}}}{4S^2} \tag{6}$$

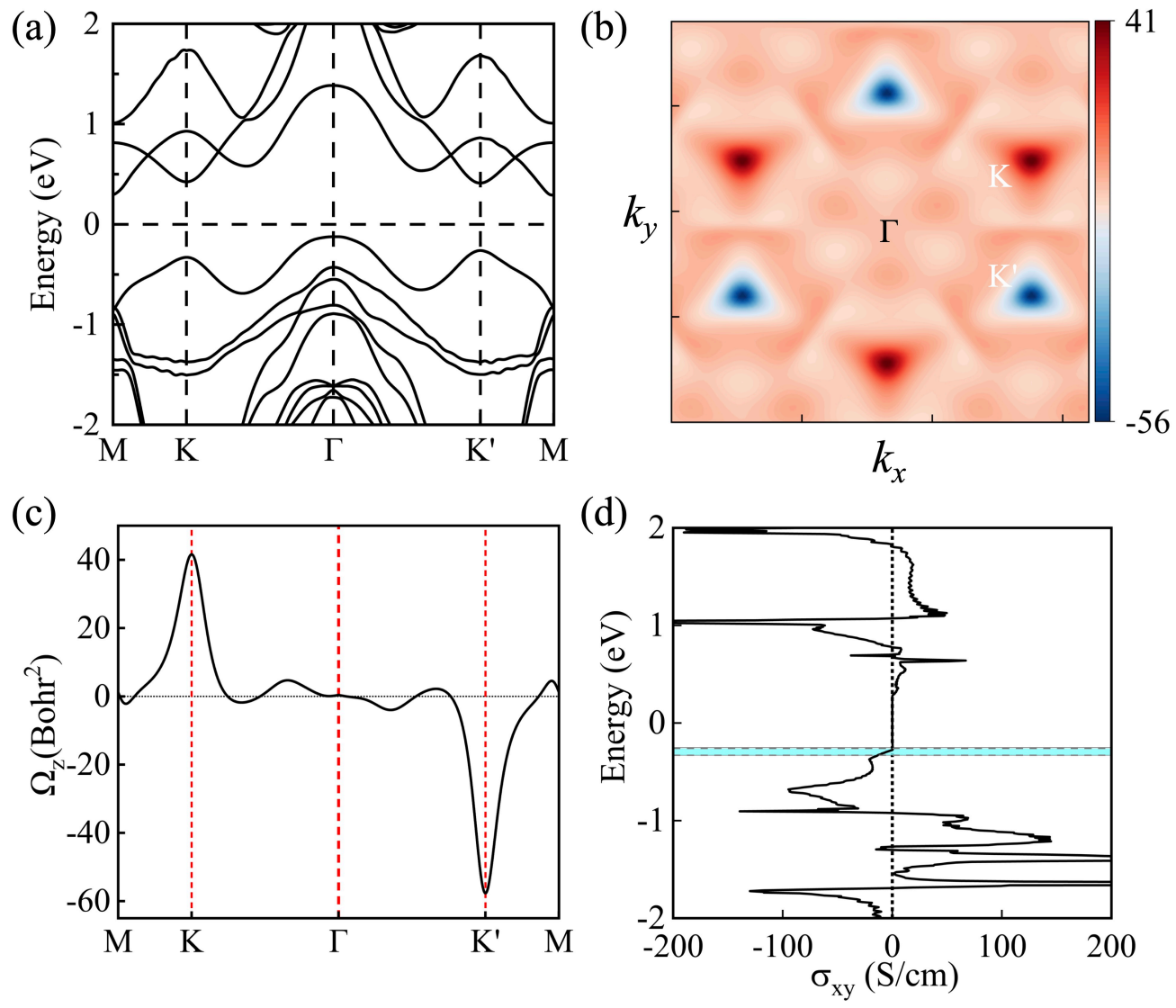

FIG. 3. (a) The Band structure of Janus 2H-VSeS monolayer calculated by MLWFs. Calculated Berry curvature of Janus 2H-VSeS monolayer (b) over 2D Brillouin zone and (c) along high symmetry lines. (d) Calculated anomalous Hall conductivity of Janus 2H-VSeS monolayer. The two parallel dashed lines indicate the two valley extremes.

where $S$ ($S$=1/2) is the spin operator, and the $\alpha$ and $z$ are the directions along the in-plane [100] and the out-of-plane [001] axis, respectively. Usually, the sub-nearest neighbor V-V interactions are negligible in this system. Thus, only the nearest V-V interactions are considered in the calculation. The calculated exchanged parameter $J$ and the magnetic ion anisotropy parameter $D$ are 64 meV and $-1.22$ meV, respectively. Figs. 2(d)-(f) illustrates the dependence of magnetic moment, magnetic susceptibility, and specific heat capacity on temperature for the Janus 2H-VSeS monolayer. The simulated $T_c$ is as high as 278 K, which is close to the room temperature (300 K).

In the case of inversion symmetry breaking in the Janus 2H-VSeS monolayer, the charge carriers also acquire a valley-contrast Berry curvature. [44] The z-component of Berry curvature $\Omega_n^z(k)$ of the occupied state band according to the Kubo formula is calculated and can be expressed as: [45, 46]

$$\Omega_n^z(k) = -\sum_{n \neq m} \frac{2\,\mathrm{Im}\,\langle \Psi_{nk} | v_x | \Psi_{mk} \rangle \langle \Psi_{mk} | v_y | \Psi_{nk} \rangle}{(E_m - E_n)^2} \qquad (7)$$

where $|\Psi_{nk}\rangle$ is a calculated wave function of the Bloch states at k points with the energy eigenvalue $E_n$, and $\upsilon_{x(y)}$ are velocity operators along $x$ and $y$ directions, respectively. To ensure the accuracy of Wannier basis functions, the tightly bound band structure is calculated using MLWFs, as shown in Fig. 3(a). The band structure is in high agreement with the DFT results [Fig. 1(e)], indicating that the resulting Wannier basis functions are sufficiently localized to ensure the precision and accuracy of the calculation. Fig. 3(b)-(c) shows the calculated results of Berry curvature along the high symmetry line and Brillouin zone, respectively. In the intrinsic ferromagnetic field of Janus 2H-VSeS monolayer, the Berry curvature has the opposite signs at the K and K′ points with different absolute magnitudes. Meanwhile, the integration of Berry curvature in the Brillouin zone gives the contribution to the anomalous Hall conductivity, which can be expressed as:

$$\sigma_{xy} = -\frac{e^2}{\hbar} \int_{BZ} \frac{d^2k}{(2\pi)^2} \Omega_n^z(k) \qquad (8)$$

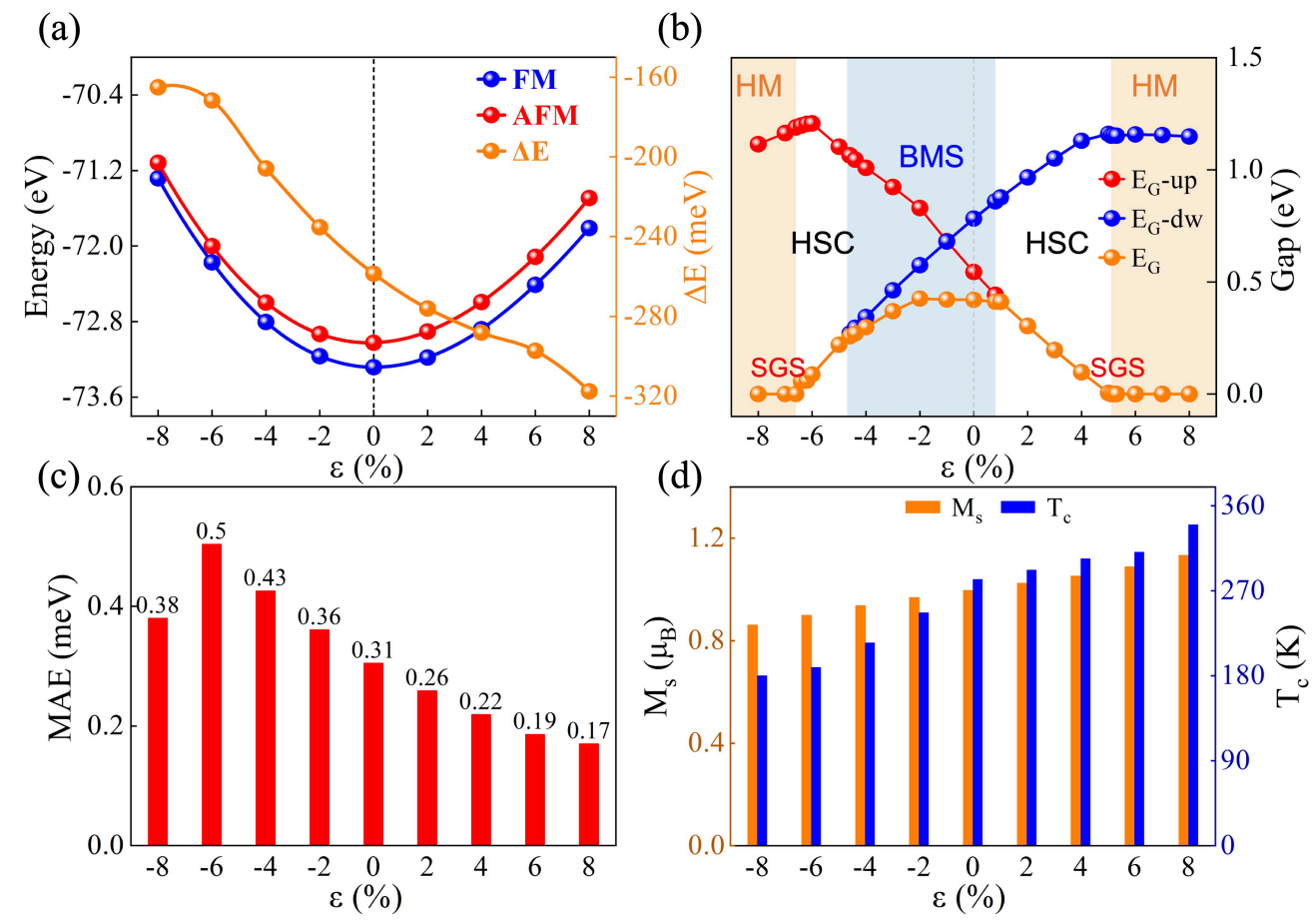

FIG. 4. The dependence of (a) ferromagnetic stability energy ΔE and (b) the band gap ($E_G$) on the strains $\varepsilon$. The red, blue, and green lines indicate $E_G$-up($\alpha$), $E_G$-down($\beta$), and $E_G$, respectively. (c) The total MAE for the Janus 2H-VSeS monolayer under various strains. (d) The magnetic moment of V atom and $T_c$ for the Janus 2H-VSeS monolayers under various strains.

Due to the breaking of the time-reversal symmetry, a non-zero anomalous Hall conductivity calculated by WannierTools appears, [47] as shown in Fig. 3(d). Since the Hall currents from the K and K′ valleys do not disappear completely, a net charge current will be generated. Due to the presence of electric field $E$, the Berry curvature drives an anomalous transverse velocity [48]:

$$v = -\frac{e}{\hbar} E \times \Omega_n(k) \qquad (9)$$

This is an intrinsic contribution to the anomalous Hall effect. In our system, the charge carriers in the K and K′ valleys have opposite transverse velocities due to the opposite sign of Berry curvature. As a result, the valley currents are generated. On the other hand, since the valley current and spin current remain constant under time reversal, valley Hall and spin Hall effects can appear in systems with constant time-reversal due to the broken reversal symmetry.

Since the CBM and VBM are contributed by in-plane $d_{x^2-y^2}/d_{xy}$ orbitals for the Janus 2H-VSeS monolayer,

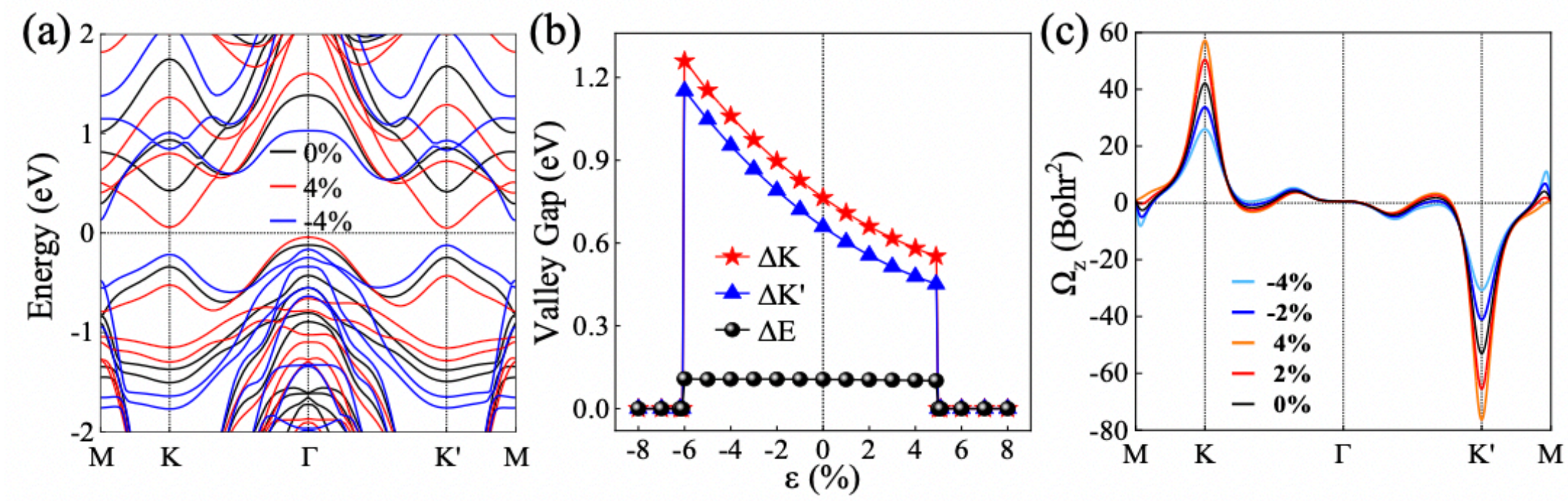

FIG. 5. (a) Band structures of Janus 2H-VSeS monolayers under various strains (±4%) with SOC. (b) Dependence of ΔK, ΔK′ and valley splitting ΔE on the strain. (c) The Berry curvature of Janus 2H-VSeS monolayers at the strains of −4%, −2%, 0%, 2%, and 4%, respectively.

it is predicted that the magnetic and electronic properties are strongly related to in-plane biaxial strain. To estimate the strain dependence of electronic properties, the in-plane biaxial strain is applied by the relation of $\varepsilon$=[(a-$a_0$)/$a_0$] ×100%, where a and $a_0$ present the lattice constants of strained and unstrained Janus 2H-VSeS monolayers. As shown in Fig. 4(a), the total energy of FM and AFM states show monotonically asymptotic change with increasing the tensile strains ($\varepsilon$=2, 4, 6, 8%) and the compressive strains ($\varepsilon$=−2, −4, −6, −8%). The ferromagnetic stability energy ΔE increases monotonically with increasing the strains from −8% (165 meV) to 8% (318 meV). The distance between V atoms becomes smaller with continuously increasing compressive strain, which leads to an increase in super-exchange interactions. However, the Janus 2H-VSeS monolayer consistently maintains the FM ground state under the various strains from −8% to 8%. It is clear from Fig. 4(b) that the in-plane biaxial strain can effectively adjust the electronic structures of Janus 2H-VSeS monolayer from the BMS character to Half-semiconductors (HSC), Spin gapless semiconductors (SGS), Half metallic (HM) characters. The unstrained Janus 2H-VSeS monolayer exhibits clear BMS character with an indirect band gap ($E_G$) of 0.423 eV, while the corresponding $E_G$-$\alpha$ and $E_G$-$\beta$ values are 0.545 eV and 0.783 eV, respectively, as shown in Fig. 4(b). At the small strain range (−4.7%<$\varepsilon$<1.0%), the Janus 2H-VSeS monolayer still maintains the BMS character. It is noted that the monotonicity of $E_G$-$\alpha$ and $E_G$-$\beta$ is very obvious at the strain range (−6.6%<$\varepsilon$<5.0%). In order to effectively discuss the change of spin-$\alpha$ and spin-$\beta$ energy bands, the difference of band edge between the spin-$\alpha$ and spin-$\beta$ electrons at the M and K points can be defined as $\Delta E_{CBM-M-K}$=$E_{CBM-K-\alpha}$−$E_{CBM-M-\beta}$. As the tensile strain increases, the $E_G$ and $E_G$-$\alpha$ values show a monotonic linear decrease, while $E_G$-$\beta$ values show a monotonic linear increase. For $\varepsilon$=0%, the $\Delta E_{CBM-M-K}$ and $E_{VBM-\Gamma-\alpha}$ values are 0.1244 and −0.2382 eV, respectively. As the strain increases to $\varepsilon$=0.8 and 1.0%, the corresponding $\Delta E_{CBM-M-K}$ values are 0.0204 and −0.0004 eV, respectively and the corresponding $E_{VBM-\Gamma-\alpha}$ increases to −0.1816 and −0.1211 eV. Therefore, it can consider that the Janus 2H-VSeS monolayer is converted to the HSC character at $\varepsilon$=1%. As the strain continues to increase from 1% to 5.0%, the HSC character is still remained. The linear fit is used as a function to describe its relationship. The $E_G$ and $\varepsilon$ can be well fitted using the relationship: $E_G$=0.5−10.06$\varepsilon$. At the same time, the $E_G$-$\alpha$ and $E_G$-$\beta$ can be also fitted with $E_G$-$\alpha$=0.51−10.06$\varepsilon$ and $E_G$-$\beta$=0.82+6.8$\varepsilon$, respectively. Figs.S3(c)-(f) shows the spin-polarized band structures of Janus 2H-VSeS monolayer under various strains. One can see that the positions of CBM and VBM relative to the Fermi level change regularly. The $E_{VBM-\Gamma-\alpha}$ values under various strains are −0.0508 eV (2%), −0.1446 eV (3%), −0.0559 eV (4%), 0 eV (5%), 0.0057 eV (5.1%) and the $E_{CBM-K-\alpha}$ values are 0.2528 eV (2%), 0.0515 eV (3%), 0.0413 eV (4%), 0.0049 eV (5%), 0.0015 eV (5.1%). Therefore, with $\varepsilon$=5.0%, only the spin-$\alpha$ electron channel is conductive, while the spin-$\beta$ electron channel is semiconductive, presenting a half-metallicity character. Meanwhile, when the $E_{VBM}$ value equals 0 eV, the Janus 2H-VSeS monolayer exhibit the SGS character. When the tensile strain continues to increase, the $E_G$ and $E_G$-$\alpha$ values are 0 eV, resulting that the Janus 2H-VSeS monolayer still maintains HM character. Under the compressive range, the difference of band edge between the spin-$\alpha$ and spin-$\beta$ electrons at the Γ point can be defined as $\Delta E_{VBM-\Gamma}$=$E_{VBM-\Gamma-\alpha}$−$E_{VBM-\Gamma-\beta}$. At the compressive strains (−4.6%<$\varepsilon$<0%), the $\Delta E_{VBM-\Gamma}$ value decreases from 0.3621 eV ($\varepsilon$=0%) to 0.2585 eV (−1%), 0.1492 eV (−2%), 0.0297 eV (−3%), −0.0995 eV (−4%) and −0.1809 eV (−4.6%) eV. At the same time, the difference of band edge between the spin-$\alpha$ and spin-$\beta$ electrons at the Γ and K points can be defined as $\Delta E_{VBM-\Gamma-K}$ = $E_{VBM-\Gamma-\beta}$-$E_{VBM-K-\alpha}$. Obviously, the $\Delta E_{VBM-\Gamma-K}$ value increases from −0.1932 eV ($\varepsilon$=0%) to −0.1643 eV (−1%), −0.1330 eV (−2%), −0.0928 eV (−3%), −0.0451 eV (−4%) and −0.0118 eV (−4.6%). When $\varepsilon$=−4.7%, the $E_{VBM-\Gamma-\beta}$ and $E_{VBM-K-\alpha}$ values are extremely close to each other, which will serve as the transition point from the BMS to HSC characters. The relationship between the $E_G$-$\alpha$($E_G$-$\beta$) and $\varepsilon$ can be expressed as: $E_G$-$\alpha$=0.57−10.86$\varepsilon$ and $E_G$-$\beta$=0.79+11.28$\varepsilon$ (−4.6%<$\varepsilon$<1%), respectively. When the compressive strain $\varepsilon$ is larger than −4.7%, the $E_{VBM-\Gamma-\beta}$ value increases from −0.2013 eV to −0.0947 eV (−5%), −0.0354 eV (−6%), −0.0025 eV (−6.4%), 0 eV (−6.42%). Thus, $\varepsilon$=−6.42% is the transition point from the HSS to SGS characters. Similar to the case of tensile strain, with the increase of compressive strain, the Janus 2H-VSeS monolayer continues to maintain the HM character. Therefore, it can conclude that the biaxial strain can effectively adjust the Janus 2H-VSeS monolayer from the BMS to HSC, SGS, and HM characters.

The total MAE for the Janus 2H-VSeS monolayer under various strains is shown in Fig. 4(c). It is clear that the Janus 2H-VSeS monolayer still remains the IMA character under various strains. The total MAE decreases linearly as the tensile strain increases from 0% to 8%. However, under the compressive strain range, the total MAE firstly increases and reaches a maximum value of 0.5 meV at the compressive strain of −6%, then gradually decreases from −6% to 0%. Fig. S4 shows the V, Se and S atomic-layer-resolved MAE under various strains. The MAE from the S and V atoms decrease linearly as the strain increases from −8% to 8%, while that from the Se atom firstly increases to a maximum value at the strain of −6% and then gradually decreases, which is similar to the total MAE behavior. Interestingly, the V and Se/S atoms always maintain the IMA character throughout the whole strain range. The main contribution of MAE in the Janus 2H-VSeS monolayer comes from the V and Se atoms, which is consistent with the fact that MAE can be contributed by non-metallic atoms with stronger SOC. [49] Fig. S5(a)-(d) shows the V-d and Se/S-p orbital resolved MAE for the Janus 2H-VSeS monolayers under various strains, respectively. As the strain increases from −8% to 8%, the positive MAE of Janus 2H-VSeS monolayer still mainly comes from the matrix element differences ($d_{x^2-y^2}$/$d_{xy}$ orbitals) of V atom and ($p_x$/$p_y$ orbitals) of Se atom. While the matrix element difference of Se ($p_z$/$p_y$ orbitals) atom shows a transition from IMA to (perpendicular magnetic anisotropy) PMA character. The matrix element difference of S ($p_x$/$p_y$ orbitals) atom still makes rather small contribution to MAE. Hence, the matrix element differences ($p_x$/$p_y$ and pz/py orbitals) of Se atom and ($d_{x^2-y^2}$/$d_{xy}$ orbitals) of V atom dominate the MAE behavior for the Janus 2H-VSeS monolayer under various biaxial strains. Fig. 4(d) shows the magnetic moment of V atom and $T_c$ for the Janus 2H-VSeS monolayer under various strains. At the tensile strain ranges ($\varepsilon$= 2, 4, 6, and 8%), the corresponding magnetic moments of V atom are 1.024, 1.052, 1.088, and 1.133 $\mu_B$, respectively. At the compressive strain ranges ($\varepsilon$= −2, −4, −6, and −8%), the corresponding magnetic moments of V atom are 0.968, 0.936, 0.899, and 0.861 $\mu_B$, respectively. It is obvious that the magnetic moment of V atom shows a monotonical increase with increasing the strains from −8% to 8%. In fact, as the strain increases, the bond length between the V atoms increases and the ionic-bond interaction overcomes the covalent-bond interaction. As a result, the number of unpaired electrons in the V atom increases, leading to an increase in the magnetic moment. This is consistent with the reported results in the 2H-$VS_2$ and 2H-$VSe_2$ monolayers. [50] One can see from Fig. 4(d) that as the strain increases from −8% to 8%, the corresponding $T_c$ are 180 (−8%), 189 (−6%), 215 (−4%), 247 (−2%), 278 (0%), 292 (2%), 304 (4%), 311 (6%) and 340 K (8%), respectively, exhibiting a monotonical increase behavior. Apparently, this is due to the further enhancement of super-exchange interaction with the increase of strain. The dependences of the magnetic moment of V atom and specific heat capacity on the temperature under various strains are shown in Fig. S6(a)-(b), and the transition behavior is consistent with the above analysis.

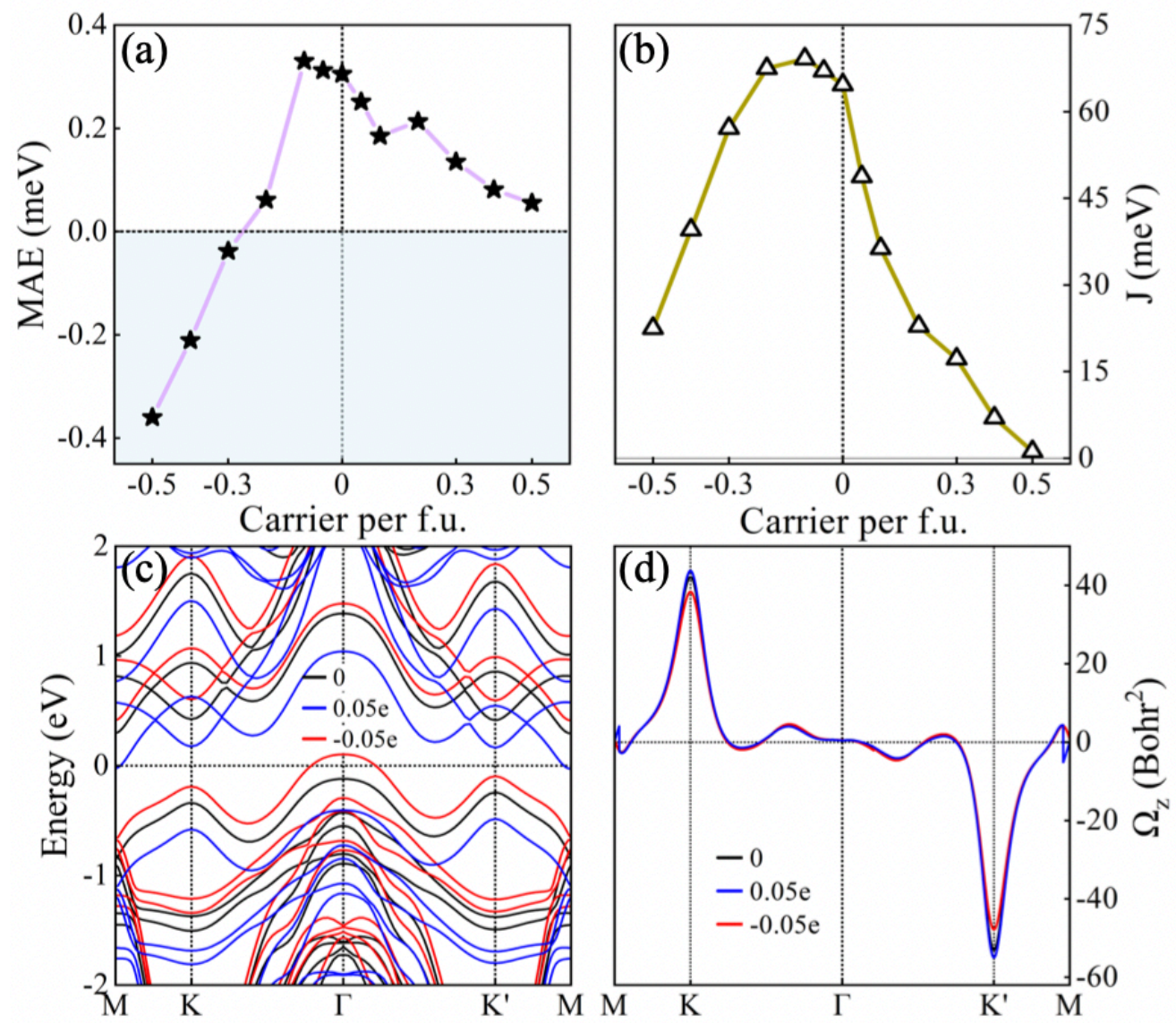

FIG. 6. The (a) MAE and (b) exchange energy $J$ for the Janus 2H-VSeS monolayer with various charge carrier doping. The (c) band structures and (d) Berry curvature for the Janus 2H-VSeS monolayer with 0.05 holes and 0.05 electrons doping, respectively. The Fermi level is set to zero.

The band structures of Janus 2H-VSeS monolayers under various biaxial strains (±4%) with considering SOC are shown in Fig. 5(a). One can see that applying strain can effectively adjust the band structure of Janus 2H-VSeS monolayer accompanied by the bands up at the compressive strain (−4%) and the bands down at the tensile one (4%) with respect to the vacuum level. [51] Therefore, it is convinced that the valley degrees of freedom can be well tuned by applying the strains. Fig. 5(b) shows the dependence of ΔK, ΔK′ and valley splitting ΔE on the biaxial strain for the Janus 2H-VSeS monolayers. When the tensile strains ($\varepsilon$=1, 2, 3, 4 and 4.9%) are applied, the corresponding ΔK values are 709, 660, 618, 580 and 552 meV, while the correspond-ing $\Delta K'$ values are 604, 556, 514, 478 and 451 meV. For the compressive strains ($\varepsilon$=−1, −2, −3, −4%, −5% and −6%), the corresponding $\Delta K$ values are 827, 897, 975, 1060, 1153 and 1260 meV, while the corresponding $\Delta K'$ values are 721, 790, 868, 953, 1048 and 1151 meV. Clearly, the $\Delta K$ and $\Delta K'$ values monotonously increase with increasing the strains from −6% to 4.9%. Surprisingly, $\Delta K$ and $\Delta K'$ values show a good quadratic dependence on the strain, namely $\Delta K = 36\varepsilon^2 - 6.02\varepsilon + 0.7636$ and $\Delta K' = 37\varepsilon^2 - 5.96\varepsilon + 0.658$. The valley splitting $\Delta E$ firstly increases and reached a maximum value of 106.8 meV at the strain of −4%, and then gradually decreases with further increasing the strain from −3% to 5%. Therefore, the valley degrees of freedom of Janus 2H-VSeS monolayer can be well modulated by the biaxial strain. The Berry curvature of Janus 2H-VSeS monolayers at the strains of −4%, −2%, 0%, 2%, and 4% is shown in Fig. 5(c). One can see that when the strain of 4% is applied, the maximum modulation of Berry curvature can reach 45%. The tensile (compressive) strain tends to increase (decrease) the Berry curvature and their magnitude difference $|\Omega_n^z(\mathrm{K}) + \Omega_n^z(\mathrm{K}')|$ . Since the AHC is defined as the integral sum of Berry curvature in the BZ, and the Berry curvature is non-zero only around the K and K′ points. The AHC is expected to be greatly enhanced under the tensile strain. In this way, one can adjust the transverse Hall voltage by applying the biaxial strains.

As is well known, charge carrier doping can be considered to an effective method to modulate the electronic structure and Fermi level of semiconductor. It has been reported that the electrical, magnetic and valley properties of 2H-$VS_2$ [52], 2H-$VSe_2$ [13], and 2H-$WSe_2$[53] monolayers can be well tuned by charge carrier doping, which can be also expected for the Janus 2H-VSeS monolayer due to the presence of unpaired electrons in the V 3d orbitals.

The MAE and exchange energy $J$ for the Janus 2H-VSeS monolayer with various charge carrier doping are shown in Figs. 6(a)-(b). It is obvious that the MAE shows a gradual decrease trend with increasing the hole or electron doping concentration. Interestingly, the electron doping does not alter the IMA character of Janus 2H-VSeS monolayer. On the contrary, the IMA character is transformed into the PMA character when the hole doping concentration is above −0.3e. On the other hand, the exchange energy $J$ still remains the positive value under various electrons (holes) doping concentrations (0-0.5e), strongly suggesting that the transition from the FM to AFM states does not occur. The $J$ firstly increases to a maximum value of 69.2 meV at the hole doping concentration of −0.1e and then gradually decreases with increasing the carrier doping concentration from −0.1e to 0.5e. Compared to the undoped case, the maximum modulation of MAE and $J$ values can reach to 82% (218%) and 98% (65%) by the electrons (holes) doping, respectively. The band structures of Janus 2H-VSeS monolayer with holes (electrons) doping is shown in Fig. 6(c). It is clear that the Fermi level shifts to the VBM (CBM) with introducing the hole (electrons) concentration of 0.05e, implying that the charge carrier doping can effectively modulate the band structures and position of the Fermi level. The lack of time-reversal symmetry and mirror symmetry makes the electronic structure of Janus 2H-VSeS monolayer more tunable compared with the nonmagnetic TMDs. The control of valley freedom by charge carrier doping is also investigated and shown in Fig. 6(d). Obviously, for the Janus 2H-VSeS monolayer doped with holes (electrons), the peak intensity of Berry curvature $\Omega_n^z(k)$ at the K and K′ points become weaker (stronger). The modulation of $\Omega_n^z(k)$ can reach 3.8% (9.5%) by doping with the electron (hole) concentration of 0.05e, respectively. Compared to the biaxial strain modulation, it is obvious that the charge carrier doping has a smaller effect on the $\Omega_n^z(k)$.

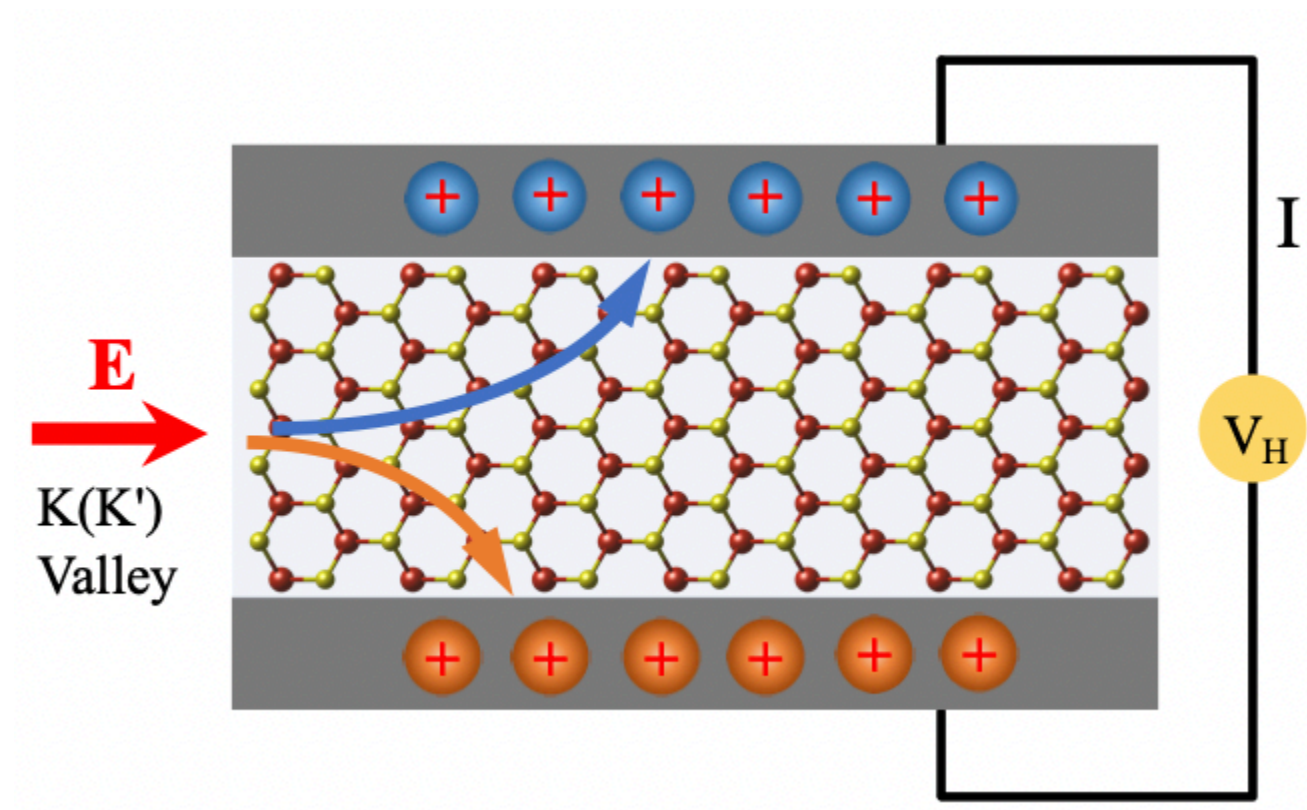

FIG. 7. Schematic of Janus 2H-VSeS monolayer for the valleytronic device. The holes are denoted by "+" symbol.

As shown in Fig. 7, a valleytronic device for the Janus 2H-VSeS monolayer can be constructed based on the calculated results of the Berry curvature $\Omega_n^z(k)$ and the anomalous Hall conductivity. With charge carrier doping, the Fermi level of Janus 2H-VSeS monolayer can be modulated between the K-valley ($\Delta K$) and K′-valley ($\Delta K'$). As an example of proper hole doping, when an in-plane electric field $E$ is applied, Janus 2H-VSeS monolayer is magnetised upwards and spin holes from the K′-valley flow upwards and the accumulated holes generate a charge Hall current that can be measured by a positive voltage. On the contrary, while reversing the Janus 2H-VSeS monolayer or modulating magnetized downward, the spin holes accumulated by flowing to the lower K-valley can be measured as a negative voltage. This device can be used as a spin filter, filtering out all carriers with spin-up and spin-down tp move laterally, which will produce a net Hall current. In summary, the Janus 2H-VSeS monolayer can be applied to anomalous Hall effect valley electronics devices and become the basis for valley electronics applications.

## IV. CONCLUSIONS

In summary, the geometry, magnetic, electronic, and valley properties of the Janus 2H-VSeS monolayer are investigated in detail using first-principles calculations. The Janus 2H-VSeS monolayer exhibits an indirect band-gap semiconductor character with ferromagnetic Curie temperature $T_c$ of 278 K, in-plane MA, and large spontaneous valley splitting of 105 meV. Interestingly, as the strain increases from −8% to 8%, the magnetic moment of V atom and $T_c$ show a monotonous increase trend, while the valley splitting and MAE firstly increase and gradually decrease. By analyzing the 3d orbital-resolved MAE of V atoms based on second-order perturbation theory, the contributions to MAE mainly originate from the matrix element differences between the $d_{xy}$ and $d_{x^2-y^2}$ orbitals of V atoms. The electronic phase transition from BMS to HSC, SGS, and HM is also observed for the Janus 2H-VSeS monolayer. The strong SOC effect, Berry curvature, and anomalous Hall conductivity under broken inversion symmetry can also be efficiently tuned by applying strain (modulation range: 45%) and charge carrier doping (modulation range: 9.5%). Finally, a valley and spin filter device is designed based on the calculation results. The versatility of Janus 2H-VSeS monolayer with considerable MAE, high $T_c$, and large valley splitting strongly suggests its potential application in two-dimensional spintronic devices.

## ACKNOWLEDGEMENTS

This work was supported by Natural Science Foundation of Tianjin City (Grant No. 20JCYBJC16540)

[1] K. S. Novoselov, A. K. Geim, S. V. Morozov, D. Jiang, M. I. Katsnelson, I. Grigorieva, S. Dubonos, and a. Firsov, Nature **438**, 197 (2005).
[2] S. Stankovich, D. A. Dikin, G. H. Dommett, K. M. Kohlhaas, E. J. Zimney, E. A. Stach, R. D. Piner, S. T. Nguyen, and R. S. Ruoff, Nature **442**, 282 (2006).
[3] K. F. Mak, C. Lee, J. Hone, J. Shan, and T. F. Heinz, Phys. Rev. Lett. **105**, 136805 (2010).
[4] A. Splendiani, L. Sun, Y. Zhang, T. Li, J. Kim, C.-Y. Chim, G. Galli, and F. Wang, Nano Lett. **10**, 1271 (2010).
[5] T. Olsen and I. Souza, Phys. Rev. B **92**, 125146 (2015).
[6] H. Zeng, J. Dai, W. Yao, D. Xiao, and X. Cui, Nat. Nanotechnol. **7**, 490 (2012).
[7] D. Xiao, G.-B. Liu, W. Feng, X. Xu, and W. Yao, Phys. Rev. Lett. **108**, 196802 (2012).
[8] K. F. Mak, K. He, J. Shan, and T. F. Heinz, Nat. Nanotechnol. **7**, 494 (2012).
[9] Q. Zhang, S. A. Yang, W. Mi, Y. Cheng, and U. Schwingenschlögl, Adv. Mater. **28**, 959 (2016).
[10] G. Aivazian, Z. Gong, A. M. Jones, R.-L. Chu, J. Yan, D. G. Mandrus, C. Zhang, D. Cobden, W. Yao, and X. Xu, Nat. Phys. **11**, 148 (2015).
[11] X. Liang, L. Deng, F. Huang, T. Tang, C. Wang, Y. Zhu, J. Qin, Y. Zhang, B. Peng, and L. Bi, Nanoscale **9**, 9502 (2017).
[12] C. Shen, G. Wang, T. Wang, C. Xia, and J. Li, Appl. Phys. Lett. **117**, 042406 (2020).
[13] J. Liu, W.-J. Hou, C. Cheng, H.-X. Fu, J.-T. Sun, and S. Meng, J. Phys.: Condens. Matter **29**, 255501 (2017).
[14] C. Wang and Y. An, Appl. Surf. Sci. **538**, 148098 (2021).
[15] Y. Cheng, Z. Zhu, M. Tahir, and U. Schwingenschlögl, Europhys. Lett. **102**, 57001 (2013).
[16] A.-Y. Lu, H. Zhu, J. Xiao, C.-P. Chuu, Y. Han, M.-H. Chiu, C.-C. Cheng, C.-W. Yang, K.-H. Wei, and Y. Yang, Nat. Nanotechnol. **12**, 744 (2017).
[17] F. Li and Y. Li, J. Mater. Chem. C **3**, 3416 (2015).
[18] A. C. de Leon, B. J. Rodier, Q. Luo, C. M. Hemmingsen, P. Wei, K. Abbasi, R. Advincula, and E. B. Pentzer, ACS nano **11**, 7485 (2017).
[19] Y. Guo, S. Zhou, Y. Bai, and J. Zhao, Appl. Phys. Lett. **110**, 163102 (2017).
[20] J. Liang, W. Wang, H. Du, A. Hallal, K. Garcia, M. Chshiev, A. Fert, and H. Yang, Phys. Rev. B **101**, 184401 (2020).
[21] J. Yuan, Y. Yang, Y. Cai, Y. Wu, Y. Chen, X. Yan, and L. Shen, Phys. Rev. B **101**, 094420 (2020).
[22] C. Xu, J. Feng, S. Prokhorenko, Y. Nahas, H. Xiang, and L. Bellaiche, Phys. Rev. B **101**, 060404 (2020).
[23] I. Dzyaloshinsky, J. Phys. Chem. Solids **4**, 241 (1958).
[24] C. Zhang, Y. Nie, S. Sanvito, and A. Du, Nano Lett. **19**, 1366 (2019).
[25] G. Kresse and J. Furthmüller, Comput. Mater. Sci. **6**, 15 (1996).
[26] G. Kresse, J. Furthmüller, and J. Hafner, Phys. Rev. B **50**, 13181 (1994).
[27] G. Kresse and J. Furthmüller, Phys. Rev. B **54**, 11169 (1996).
[28] G. Kresse and D. Joubert, Phys. Rev. B **59**, 1758 (1999).
[29] J. P. Perdew, K. Burke, and M. Ernzerhof, Phys. Rev. Lett. **77**, 3865 (1996).
[30] P. Söderlind, O. Eriksson, B. Johansson, and J. Wills, Phys. Rev. B **50**, 7291 (1994).
[31] P. Hohenberg and W. Kohn, Phys. Rev. **136**, B864 (1964).
[32] S. Grimme, J. Comput. Chem. **25**, 1463 (2004).
[33] A. Liechtenstein, V. I. Anisimov, and J. Zaanen, Phys. Rev. B **52**, R5467 (1995).
[34] H. J. Monkhorst and J. D. Pack, Phys. Rev. B **13**, 5188 (1976).
[35] S. Grimme, J. Antony, S. Ehrlich, and H. Krieg, J. Chem. Phys. **132**, 154104 (2010).
[36] A. Togo and I. Tanaka, Scripta Mater. **108**, 1 (2015).
[37] N. Marzari, A. A. Mostofi, J. R. Yates, I. Souza, and D. Vanderbilt, Rev. Mod. Phys. **84**, 1419 (2012).
[38] A. A. Mostofi, J. R. Yates, Y.-S. Lee, I. Souza, D. Vanderbilt, and N. Marzari, Comput. Phys. Commun. **178**, 685 (2008).
[39] L. Liu, X. Ren, J. Xie, B. Cheng, W. Liu, T. An, H. Qin, and J. Hu, Appl. Surf. Sci. **480**, 300 (2019).

[40] L. Liu, S. Chen, Z. Lin, and X. Zhang, J. Phys. Chem. Lett. **11**, 7893 (2020).
[41] H. L. Zhuang and R. G. Hennig, Phys. Rev. B **93**, 054429 (2016).
[42] M. Esters, R. G. Hennig, and D. C. Johnson, Phys. Rev. B **96**, 235147 (2017).
[43] H. Zhang, L.-M. Liu, and W.-M. Lau, J. Mater. Chem. A **1**, 10821 (2013).
[44] D. Xiao, W. Yao, and Q. Niu, Phys. Rev. Lett. **99**, 236809 (2007).
[45] D. J. Thouless, M. Kohmoto, M. P. Nightingale, and M. den Nijs, Phys. Rev. Lett. **49**, 405 (1982).
[46] Y. Yao, L. Kleinman, A. MacDonald, J. Sinova, T. Jungwirth, D.-s. Wang, E. Wang, and Q. Niu, Phys. Rev. Lett. **92**, 037204 (2004).
[47] Q. Wu, S. Zhang, H.-F. Song, M. Troyer, and A. A. Soluyanov, Comput. Phys. Commun. **224**, 405 (2018).
[48] D. Xiao, M.-C. Chang, and Q. Niu, Rev. Mod. Phys. **82**, 1959 (2010).
[49] B. Yang, X. Zhang, H. Yang, X. Han, and Y. Yan, J. Phys. Chem. C **123**, 691 (2018).
[50] Y. Ma, Y. Dai, M. Guo, C. Niu, Y. Zhu, and B. Huang, ACS nano **6**, 1695 (2012).
[51] D. Zhang and S. Dong, Prog. Nat. Sci. **29**, 277 (2019).
[52] N. Luo, C. Si, and W. Duan, Phys. Rev. B **95**, 205432 (2017).
[53] R. Mukherjee, H. Chuang, M. Koehler, N. Combs, A. Patchen, Z. Zhou, and D. Mandrus, Phys. Rev. Appl. **7**, 034011 (2017).

# — Supplementary Materials —

## Supplementary Materials: Controlled Curie temperature, magneto-crystal anisotropy, and valley polarization in 2D ferromagnetic Janus 2H-VSeS monolayer

Cunquan Li[1], Yukai An[1,*]

[1] *Key Laboratory of Display Materials and Photoelectric Devices, Ministry of Education, Tianjin Key Laboratory for Photoelectric Materials and Devices, National Demonstration Center for Experimental Function Materials Education, School of Material Science and Engineering, Tianjin University of Technology, Tianjin, 300384, China*

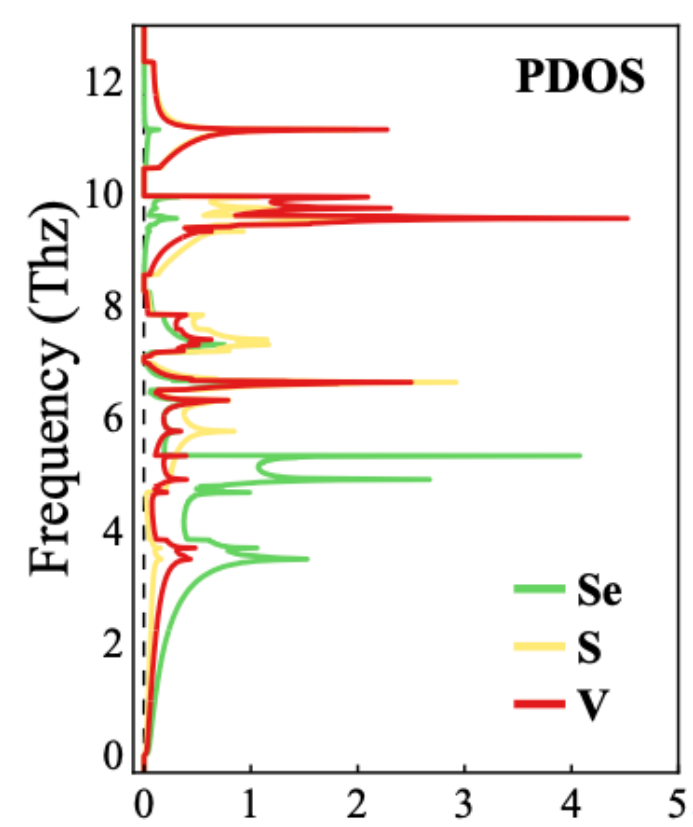

FIG. S1. Phonon density of states for the Janus 2H-VSeS monolayer. The green, yellow, and red lines present Se, S and V atom-projected phonon DOSs, respectively.

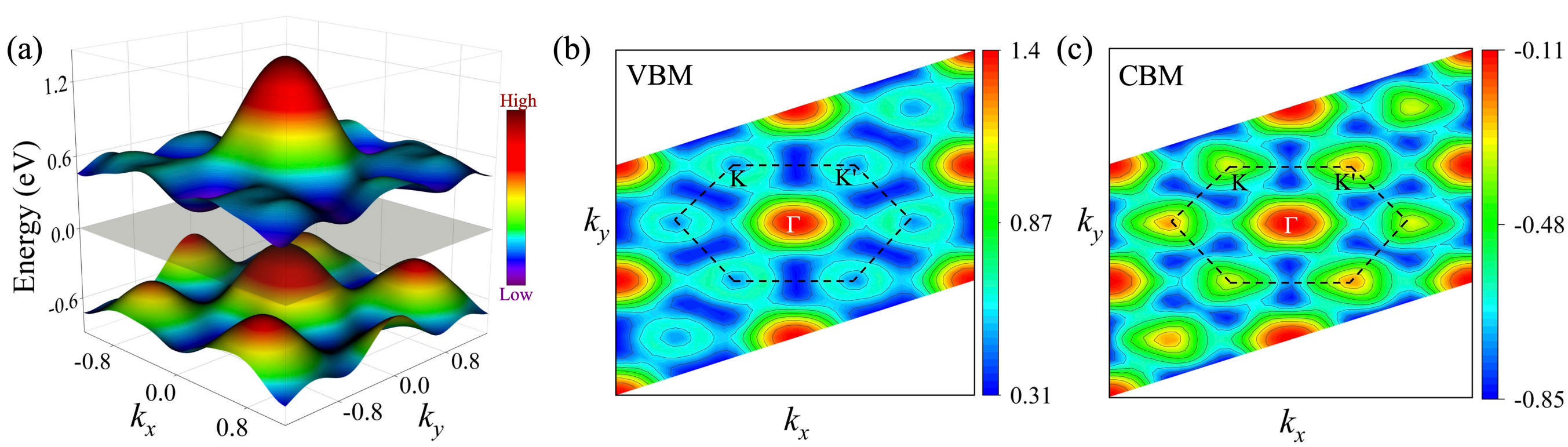

FIG. S2. (a) 3D band structure of Janus 2H-VSeS monolayer with considering SOC. Different colors indicate different equivalence surfaces. (b) 2D Projected band structure of (b) valence band and (c) conduction band at the $k_x k_y$-plane.

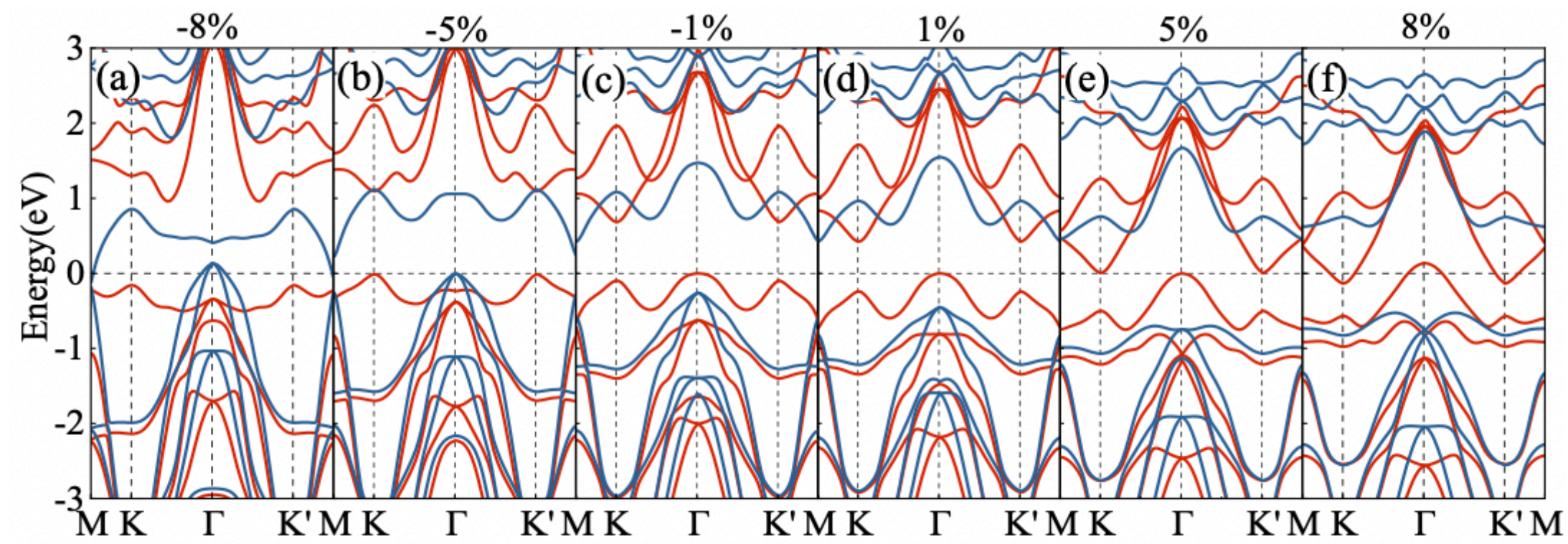

FIG. S3. The spin-polarized band structures of Janus 2H-VSeS monolayer at the strains of (a) −8%, (b) −5%, (c) −1%, (d) 1%, (e) 5% and (f) 8%, respectively.

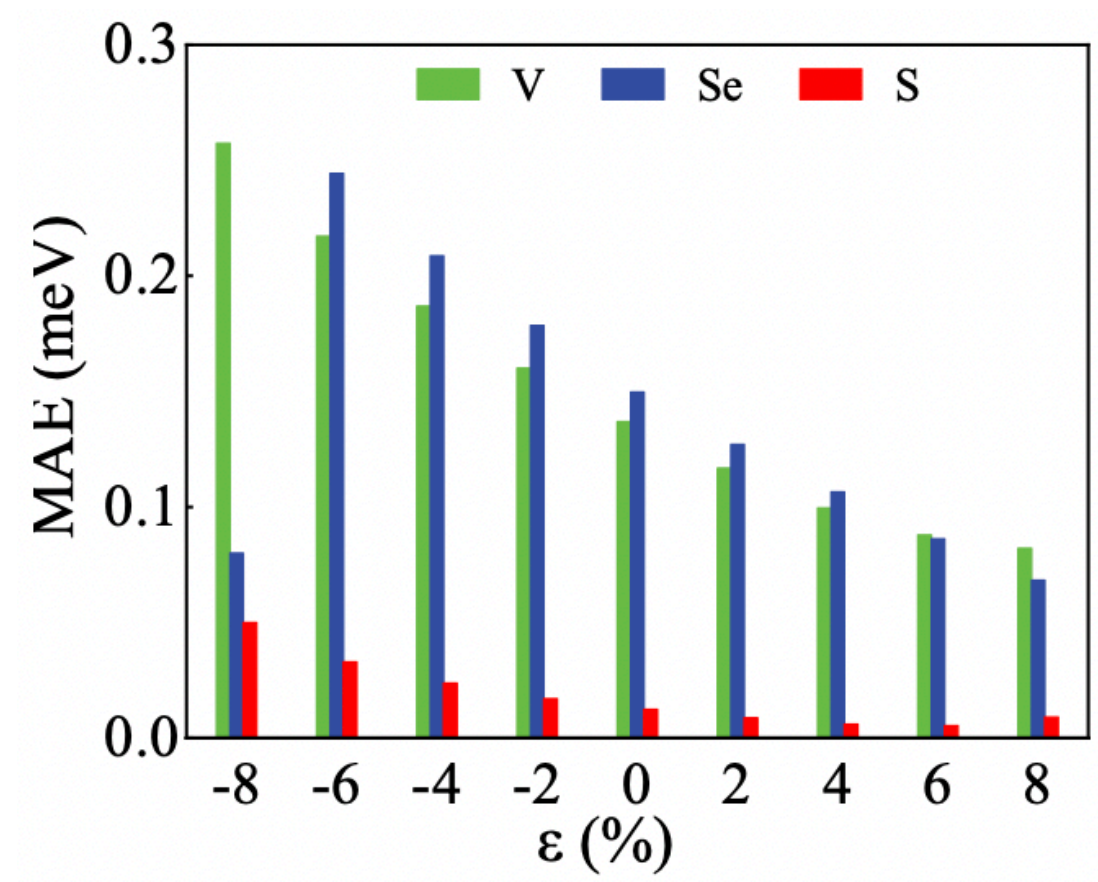

FIG. S4. The V, Se and S atomic-layer-resolved MAE for the Janus 2H-VSeS monolayer under various strains.

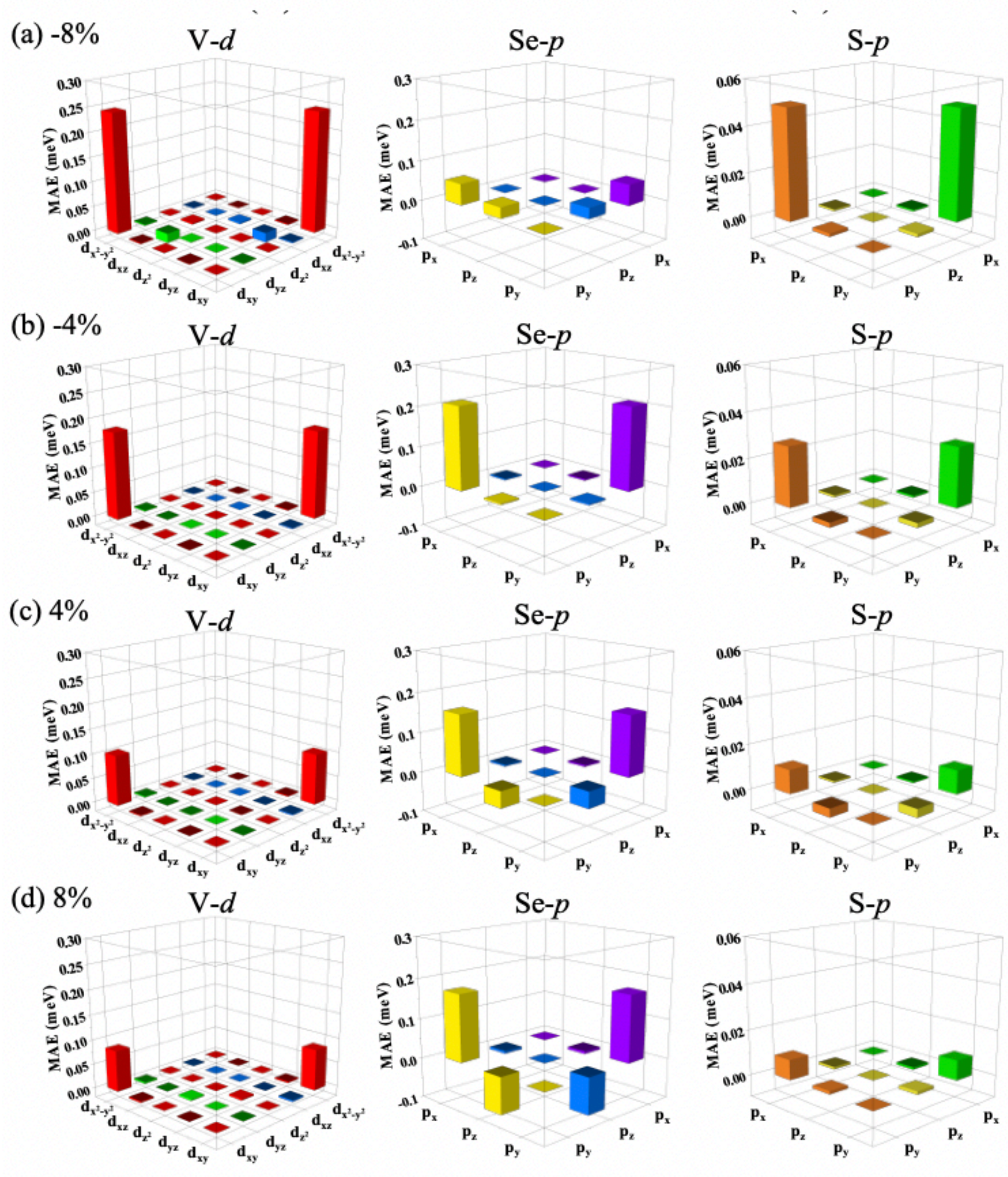

FIG. S5. The V-d and Se(S)-p orbital resolved MAE for the Janus 2H-VSeS monolayer at the strains of (a) −8%, (b) −4%, (c) 4% and (d) 8%.

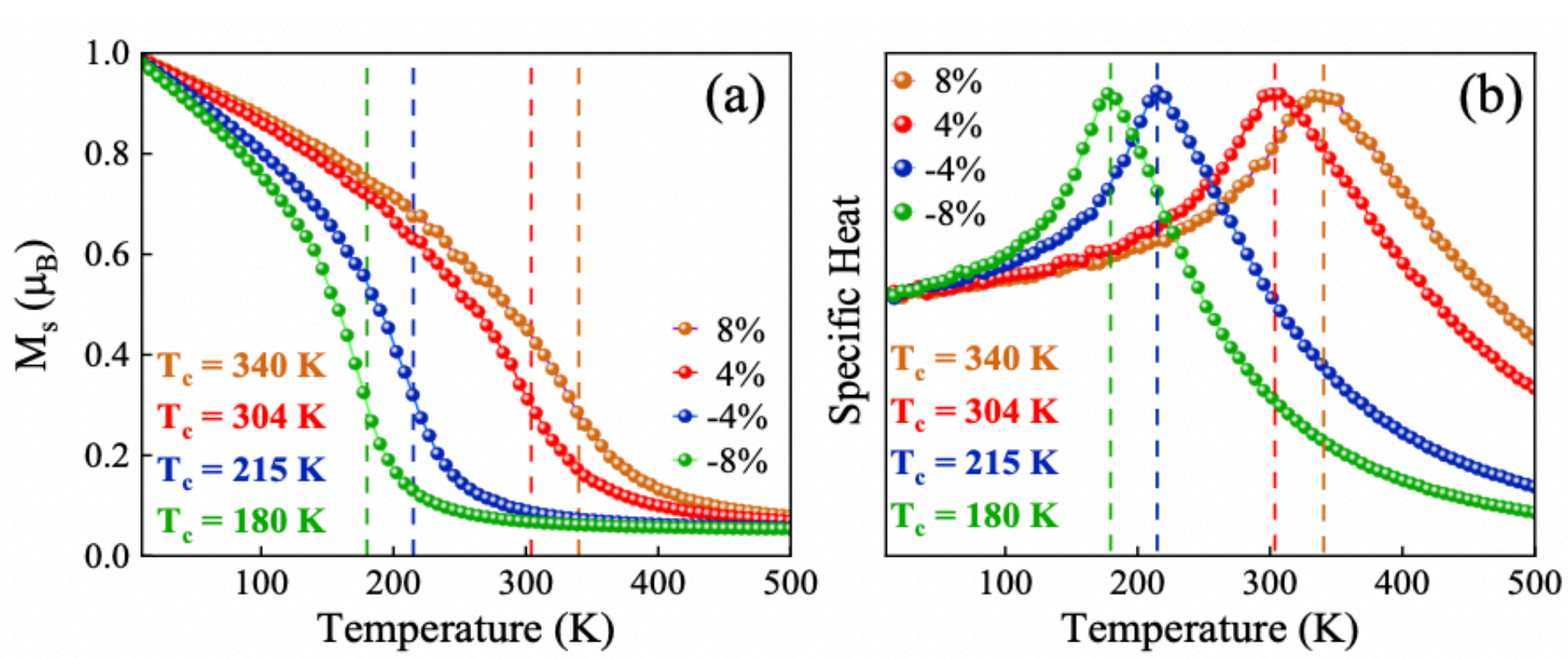

FIG. S6. Dependence of (a) magnetic moment and (b) specific heat capacity on the temperature for the Janus 2H-VSeS monolayer at the strains of −8%, −4%, 4% and 8%.